\documentclass[nonacm,10pt]{acmart}

\usepackage{booktabs} 
\usepackage{graphicx}
\usepackage{balance}  
\usepackage{todo}
\usepackage{algpseudocode}
\usepackage{algorithm}
\usepackage{hyperref}
\usepackage{cleveref}
\usepackage{enumitem}
\usepackage{multirow}

\DeclareMathOperator*{\argmax}{argmax} 

\newcommand{\squishitemize}{
 \begin{list}{$\bullet$}
  { \setlength{\itemsep}{0pt}
     \setlength{\parsep}{3pt}
     \setlength{\topsep}{3pt}
     \setlength{\leftmargin}{1.95em}
     \setlength{\labelwidth}{1.5em}
     \setlength{\labelsep}{0.5em} } }

\newcounter{Lcount}
\newcommand{\squishlist}{
    \begin{list}{\arabic{Lcount}. }
   { \usecounter{Lcount}
        \setlength{\itemsep}{0pt}
        \setlength{\parsep}{3pt}
        \setlength{\topsep}{3pt}
        \setlength{\partopsep}{0pt}
        \setlength{\leftmargin}{2em}
        \setlength{\labelwidth}{1.5em}
        \setlength{\labelsep}{0.5em} } }

\newcommand{\squishend}{\end{list}}
\newcommand{\Peloton}{Peloton}
\setcopyright{none}

\acmDOI{10.475/123_4}

\acmConference[]{}{}{}
\acmYear{}
\copyrightyear{}

\acmArticle{}
\acmPrice{}

\begin{document}
\title{Scheduling OLTP Transactions via Machine Learning}

\author{Yangjun Sheng}
\affiliation{%
  \institution{Carnegie Mellon University}
  \city{Pittsburgh}
  \state{PA}
}
\email{yangjuns@andrew.cmu.edu}

\author{Anthony Tomasic}
\affiliation{%
  \institution{Carnegie Mellon University}
  \city{Pittsburgh}
  \state{PA}
}
\email{anthony.tomasic@gmail.com}

\author{Tieying Zhang}
\affiliation{%
  \institution{Alibaba Group}
}
\email{tieying.zhang@alibaba-inc.com}

\author{Andrew Pavlo}
\affiliation{%
  \institution{Carnegie Mellon University}
  \city{Pittsburgh}
  \state{PA}
}
\email{pavlo@cs.cmu.edu}
\renewcommand{\shortauthors}{Y. Sheng, et al.}

\begin{abstract}
Current main memory database system architectures are still challenged by high contention workloads and this challenge will continue to grow as the number of cores in processors continues to increase~\cite{Yu:2014:SAE:2735508.2735511}. These systems 
schedule transactions randomly across cores to maximize concurrency and to produce a uniform load across cores. Scheduling never considers potential conflicts. Performance could be improved if scheduling balanced between concurrency to maximize throughput and scheduling transactions linearly to avoid conflicts.
In this paper, we present the  design of several intelligent transaction scheduling algorithms that consider both potential transaction conflicts and concurrency.  
To incorporate reasoning about transaction conflicts, we develop a supervised machine learning model that estimates the probability of conflict. This model is incorporated into several scheduling algorithms. In addition, we integrate an unsupervised machine learning algorithm into an intelligent scheduling algorithm. 
We then empirically measure the performance impact of different scheduling algorithms on OLTP and social networking workloads. Our results show that, with appropriate settings, intelligent scheduling can increase throughput by 54\% and reduce abort rate by 80\% on a 20-core machine, relative to random scheduling. In summary, the paper provides preliminary evidence that intelligent scheduling significantly improves DBMS performance.
\end{abstract}

%
%



\maketitle

\section{Introduction}


Transaction aborts are one of the main sources of performance loss in main memory OLTP systems~\cite{Yu:2014:SAE:2735508.2735511}.
Current  architectures  for  OLTP DBMS  use  random  scheduling  to  assign  transactions  to threads. Random scheduling achieves uniform load across CPU cores and keeps all cores occupied. For workloads with a high abort rate, a large portion of work done by CPU is wasted. In contrast, the abort rate drops to zero if all transactions are scheduled sequentially into one thread. No work is wasted through aborts, but the throughput drops to the performance of a single hardware thread.  Research has shown that statistical scheduling of transactions using a history can achieve low abort rate and high throughput~\cite{ZhangICDE2018} for partitionable workloads.  We propose a more systematic machine learning approach to schedule transactions that  achieves low abort rate and high throughput for both partitionable and non-partitionable workloads.

The fundamental intuitions of this paper are that (i) the probability that a transactions will abort will high probability can be modeled through machine learning, and (ii) given that an abort is predicted, scheduling can avoid the conflict without loss of response time or throughput. Several research works~\cite{Bailis:2014:VLDB,Stonebraker:2007:EAE:1325851.1325981,Thomson:2012:SIGMOD,prasaad-2018} perform exact analyses of aborts of transaction statements that are complex and not easily generalizable over different workloads or DBMS architectures. Our approach uses  machine learning algorithms to model the probability of aborts. 
Our approach is to (i) build a machine learning model that helps to group transactions that are likely to abort with each other, (ii) assign each group of transactions to a first-in-first-out (FIFO) queue, and (iii) monitor concurrency across cores and adjust for imbalances. The approach requires only a thin API to the database. 

\section{Challenges}
The first challenge is to learn a model to predict transactions aborts. This challenge involves dataset collection, feature space design and algorithm selection. The cost of learning and maintaining a model is important, but the performance of evaluation of an instance of the model, which occurs with every new transaction, is critical. Since our target is main-memory DBMS, we limit our design space to instance evaluations that are at most linear with respect to the length of a transaction statement. Since main-memory databases are fast, we have little time to make decisions because the competitor, randomize queuing of transactions, takes O(1) time.

The second challenge is to model when transaction conflict will occur. Modeling when an abort will occur is not obvious because transactions processing is complex and data dependent. Whether a transaction will abort depends on various static factors, the concurrency control protocol, the granularity of object representation (typically a tuple) and various dynamic factors, i.e.~the concurrent order of execution. The feature vector of the model must capture the relevant information for a model to accurately predict aborts.

The third challenge derives from the need to cluster transactions together in a way that increases throughput by avoiding aborts. Note that aborts in main-memory databases are cheap: reducing the number of aborts provides less benefit than in a disk-based DBMS.

The fourth challenge is the cost of running schedulers and training models. Given an efficient and accurate model that predicts aborts, leveraging this information in a scheduler is not straightforward since eliminating all aborts is not necessarily the optimal policy: the scheduler must simultaneously balance (potential) aborts with a dynamically changing degree of parallelism  to maximize throughput.

Finally, the approach must generalize over workloads that are not easily partitionable. We contend that, properly designed, an intelligent scheduler will reduce aborts and increase throughput relative to random scheduling. 

The contributions of this paper are the following:
\begin{itemize}
    \item We demonstrate an efficient machine learning model that predicts aborts for a pair of transactions with high accuracy.
    \item We present transaction scheduling methods that are largely independent of the internal logic of the DBMS and thus the methods can be easily adopted to any DBMS. The method has a low cost of engineering effort.
    \item We present several scheduling algorithms based on supervised and unsupervised machine learning algorithms that result in lower abort rate and higher throughput than the state of the art (random scheduling). 
    \item We show that intelligent scheduling algorithms  achieve higher throughput even when the workload is not partitionable.
    \item Our implementation of the best performing intelligent scheduler requires about 500 lines of additional code.
\end{itemize}

In summary, we report preliminary results on a new approach for intelligent scheduling in a DBMS. In Section 3, we explain the system architecture and environment. In Section 4 and 5, we  describe how to apply both supervised and unsupervised machine learning algorithms to the transaction scheduling problem. Section 6 describes our experimental framework on main-memory databases.  Section 7 describes experimental results over three workloads. Two workloads are focused on OLTP (TPC-C~\cite{TPCC}, TATP~\cite{TATP}), and one workload on social networking websites (Epinions~\cite{Epinions}). The latter benchmark workload is not easily partitionable. Section 8 discusses related work and Section 9 concludes the paper.

\section{System Architecture Overview}
\begin{figure}[t!]
    \centering
    \includegraphics[width=8cm,clip]{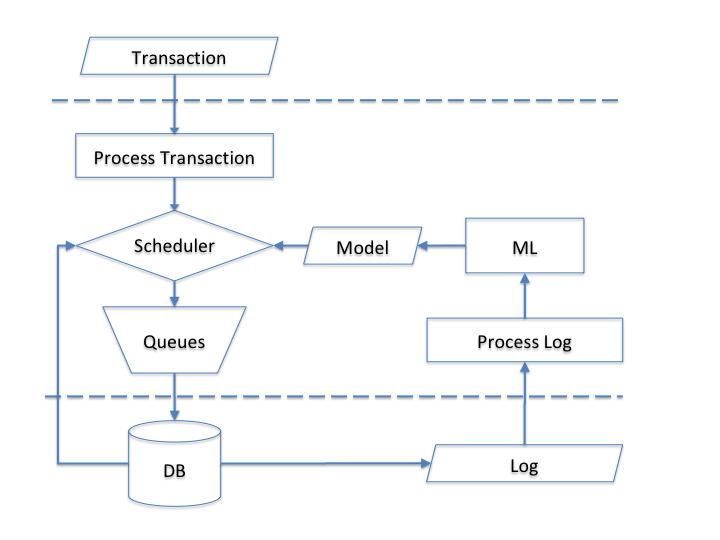}
    \caption{
        \textbf{Functional Architecture} -- 
        The Transaction component represents an incoming transaction.  The Process Transaction component converts the transaction into its feature representation. The Scheduler component chooses the queue to place the incoming transaction. The Queues component represents the set of run queues. The DB component is a main-memory database system. The Log component collects transaction execution results for model training purposes. The Process Log component converts the log into its feature representation for training. The ML component trains the processed log to produce the Model. The dotted lines separate conventional components from our scheduler. The DB component also provides real-time transaction response time information to the scheduler.
    }
\label{fig:arch}
\end{figure}

The overall environment 
(Figure \ref{fig:arch}) 
consists of three layers -- the incoming stream of transaction requests, our scheduler, and a main-memory database system. 
The API between the system and the database is simple, consisting of three functions: (i) the ability to capture incoming transactions, (ii) the ability to queue a transaction in a specific run queue of the database, (iii) the ability to log the SQL of two events that occur during transaction processing: a transaction abort and a transaction commit, and (iv) the ability to report real-time transaction response times. This environment and API implies our approach can be ``bolted on'' to any DBMS with little effort.
Our system does however require that for a (non user) aborted transaction, the system can identify at least one transaction that 'caused' the abort. Typically the causing transaction holds a lock on a row or modified the row during the aborted transaction's execution.

\section{Supervised Machine Learning}
Transactions aborts are a main source of degeneration of performance in database systems. The task in this section is to design a machine learning model that detects aborts. A naive approach would be to use supervised machine learning to learn a classifier to partition transactions into two classes, Abort and Commit by simply training over log of transactions execution results. However, this approach omits the crucial fact that a transaction execution result depends on other concurrent transactions. Our approach is to learn a classifier that classifies two transactions, $T_A$ and $T_B$, into Abort if they will abort each other when run concurrently (with high probability) or Commit if they do not conflict. 

\subsection{Training Data}
 Each transaction indirectly describes information about which tuples, attributes and tables it will read and write in the database.  To access this information, some works use the read/write set of a transaction, but this set of information is large and dynamically changing.
 Instead, the string representation of the SQL statements of a transaction are coded into a feature space. The SQL statements are a lightweight feature representation of the transaction.   
\subsubsection{Features}
As described above, the primary goal of the model is to check if two transactions will conflict with each other. Therefore, the features should be derived from both transactions, $T_1$ and $T_2$. In particular, we want to define a function \[trans(T_1, T_2) \rightarrow vector\] to transform two transactions into a feature vector and train our model to classify this vector into Abort or Commit. 

For example, in TPC-C, a transaction $T$ might want to read rows where warehouse id is equal to $10$. Then $T$ has a reference \texttt{W\_ID=10} and the string \texttt{W\_ID=10} is considered as a feature of $T$. More precisely, any instance of \textit{attribute operator value} in a \textbf{WHERE} clause of a \textbf{SELECT}, \textbf{DELETE} or \textbf{UPDATE} statement is a feature. All values of an \textbf{INSERT} are also features. We do not distinguish between reads and writes. If $T$ both read and write rows \texttt{W\_ID=10}, it has only one such string. The function $Features(string) \rightarrow F$ maps from a SQL string to a set of features $F$. Note that any particular transaction produces only a few features.

To produce a compact representation with a fixed size independent of the size of the transaction, we apply feature hashing, a fast and space-efficient way of converting an arbitrary number of features into indices in a fix-size vector. The function 
$Hash(T)$ 
constructs the union of the set of the features $F$ of  the statements of the transaction and generates a binary vector of fixed length $k$. Given two transactions $T_1$ and $T_2$, we now have two vectors $Hash(Features(T_1)) = V_1$ and $Hash(Features(T_2)) = V_2$. Let $V_3 = V_1 \& V_2$ be the binary logical AND of $V_1$ and $V_2$. The final feature vector is the concatenation of these three vectors, \[trans(T_1, T_2) = V_1|V_2|V_3\]
The vector $V_3$ encodes tuples, columns or tables that will potentially be touched by both transactions $T_1$ and $T_2$. In our experiments our feature vectors are 1k bits in length.

\subsubsection{Canonical Features}
Attribute names that appear in a schema are arbitrary, independent of the underlying domain concept. For instance, in TPC-C benchmark, the column \texttt{W\_ID} in \texttt{WAREHOUSE} table and \texttt{D\_W\_ID} in \texttt{DISTRICT} table both represent warehouses in this database. However, their string representations in SQL are not the same (\texttt{W\_ID=1} and \texttt{D\_W\_ID=1}). While they  express  the same  entity, based on the hashing function described in previous section, these two strings hash to different indices in the feature vector (baring collision). Research has shown that canonical features are more favorable than literal strings for representing transactions~\cite{ZhangICDE2018}. 
Therefore, we adopt canonical features converting string representations of attributes to a canonical version. 

Suppose a TPC-C $New Order$ transaction contains following SQL statements:
\begin{verbatim}
 SELECT W_NAME, W_STREET_1, W_STREET_2, W_CITY, 
 FROM WAREHOUSE WHERE W_ID = 10
 ...
 SELECT D_NAME, D_STREET_1, D_STREET_2, D_CITY
 FROM DISTRICT WHERE D_W_ID = 10 AND D_ID = 4
 ...
\end{verbatim}
Then the feature representation for these two SQL statements is: 
\begin{verbatim}
W_ID = 10,..., W_ID = 10, D_ID = 4, ...
\end{verbatim}
That is, each of these three strings is hashed and the corresponding bit is turned on in the feature vector. This representation discards most of the meaning of the SQL query and concentrates on representing only the read/write footprint of the transaction.  The representation is invariant to the ordering of expressions in conjunctions and disjunctions. More complex SQL expressions, such as range expressions, are reserved for future work.

\subsubsection{Training}
Training data is gathered by observing the log while the system is running. Every time a transaction commits or aborts, the event is logged as one of two possible abstract triples:
\begin{itemize}
    \item (Aborted Transaction, Conflicting Transaction, label \verb|abort|)
    \item (Committed Transaction, Any Concurrent Transaction, label \verb|commit|)
\end{itemize}
Sampling the log under random scheduling generates a distribution of transactions and abort/commit independently of our subsequent scheduling decisions. Concretely, if $T_1$ is aborted by $T_2$, then we log the pair $(trans(T_1, T_2), 1)$ into our log where $1$ indicates an abort. To obtain vectors in the commit set, we used the following approach: when $T_1$ commits, we randomly pick a transaction, $T_2$, that is running currently with $T_1$ and add $(trans(T_1, T2), 0)$ into our log. The log serves directly as the training and test set for choosing a learning algorithm.

Note that only one conflicting transaction is chosen to record in the log. However, since the choice is random, the log implicitly records a random sample of the distribution of conflicting transactions, exactly in proportion to their occurrence in the workload. 

Moreover this feature encoding intentionally ignores many details of the transaction process system: the type of concurrency control regime, the level of conflict detection (partition, tuple, field, object, page), etc. All these issues are treated as a ``black box'' by the machine learning model. The model learns which features are relevant to modeling abort probabilities as given by the log. Thus, our approach is applicable to a wide range of systems.

\subsubsection{Model Evaluation}

We will assume that the hypothesis space is linear and defer non-linear models for future work. Our learning algorithm needs to be cheap to train and very fast to evaluate on a pair of transactions. Logistic regression is both cheap to train and can evaluate an example in $O(n)$ time where $n$ is the number of 1 bits in the feature vector, using a sparse representation. Example evaluation time is proportional to the sum of the string lengths of the two transactions. In practice $n$ is small (less than 20) and the evaluation consists of multiplication of each 1 feature vector bit by a scalar and summing to produce a final score. Another advantage is that the score for logistic regression can be interpreted as the probability that the transaction pair will produce an abort. The trained model generates parameters for a function \[M(T_1, T_2) \rightarrow P(commit/abort) \] that predicts abort probabilities for any transaction pair $(T_1, T_2)$.

\subsubsection{Model Accuracy}
With a large set of transaction abort instances, we use 4-fold cross-validation to learn and evaluate the accuracy of the models. We set the training size to include 10,000 training data points for each benchmark.   Logistic regression (Table~\ref{table:model_accuracy}) performs well for all three benchmarks. 
\begin{table}[h]
\begin{center}
\begin{tabular}{c|c|c|c}
\multicolumn{3}{c}{\textbf{Model Accuracy}} \\
Benchmark           & Training Size & Accuracy & Majority\\ \hline \hline
\verb TATP &  10,000     & 0.966 & 0.500\\
\verb TPC-C & 10,000     & 0.985 & 0.500\\
\verb Epinions     & 10,000     &  0.951 & 0.500\\
\multicolumn{3}{c}{ } \\ 
\end{tabular}
\caption{Logistic Regression Accuracy. The Accuracy column reports the accuracy of the logistic regression classifier. For comparison, the Majority column reports the accuracy of a classify that chooses the majority class of the training data. }
\label{table:model_accuracy}
\end{center}
\end{table}

\subsection{Scheduling Algorithms}
Our task in this section is to design an algorithm using $M$ that assigns a transaction, $T_{new}$, into a queue in a way that avoids aborts and increases throughput compared to random scheduling.  

Consider a transaction $T_k$ already queued and waiting to execute. Suppose $M(T_{new}, T_k)$ has low abort probability. A low abort probabilities does not make it clear which queue to assign $T_{new}$ since some other concurrent transaction may have a high abort probability. 
Suppose $M(T_{new}, T_k)$ predicts a high abort probability. Then if these two transactions run concurrently, with high probability at least one abort will occur, perhaps two (depending on the details of the concurrency control system). Assigning transaction $T_{new}$ into a different queue than $T_k$ may succeed as long as the database does not try to execute them at the same time. Concurrent execution depends in this case on there being the same amount of work in front of each queued transaction, so that they both arrive to the front of the queue at the same time, and are processed concurrently. 
However, predicting the total work in a queue is difficult. The simplest heuristic places $T_{new}$ into the same FIFO queue after $T_k$ so they never execute concurrently.

With this heuristic in mind, the algorithm computes
\[Enqueue(T_{new},queue(\argmax_{T_i} M(T_{new}, T_i))\] 
where $queue(T)$ is a function that returns the queue of transaction $T$. That is, for a new transaction $T_{new}$, search for  $T_{high}$ in some queue with the  highest abort probability $M(T_{new}, T_{high})$, then assign $T_{new}$ to the same queue as $T_{high}$ (\cref{algo:search}).  
\begin{algorithm}
\caption{Assign($T_{new}$)}
\begin{algorithmic}
\State $p=-1$
\State $q=-1$
\For{$T \in \mathbb{T}$}
\If{$M(T_{new}, T) > p$}
    \State $p = M(T_{new}, T)$
    \State $q = queue(T)$
\EndIf
\EndFor
\If{$p > -1$}
    \State Assign $T_{new}$ to FIFO $q$
\Else
    \State Assign $T_{new}$ randomly
\EndIf
\end{algorithmic}
\label{algo:search}
\end{algorithm}


\subsubsection{Search Scheduling Algorithm}


Usually, more than one scheduler assigns transactions concurrently. In this case, two schedulers respective inflight transactions might conflict. Dealing with this case would require the schedulers to check if their  transactions conflict, requiring a costly coordination between schedulers. 

Finally, for efficiency purposes, if a high probability abort is discovered (above $\theta=0.5$), the search is stopped early and the current $T$ is treated as $T_{high}$. The new transaction is added to the end of $T$'s queue. To prevent bias towards one queue or another, we randomize the starting queue for the search.

We can search queues in two different ways: Breadth First Search (BFS) and Depth First Search (DFS). Extensive analysis and experimentation have shown that DFS performs poorly compared to BFS. DFS spends much more time searching for high probability transactions.  Suppose we have $n$ queues, $q_0, q_2, ... q_{n-1}$. The pseudo code for assigning $T_{new}$ using BFS scheduling (\cref{algo:bfs}) is below. The code assumes that $ID(T)$ returns the identifier of $T$ which is also a timestamp of the starting time of the transaction.

\begin{algorithm}[t]
\caption{BFSAssign($T_{new}$)}
\begin{algorithmic}
\State ${id\_set} = \{ \}$
\State ${q\_set} = \{ \}$
\State $\alpha \gets Random(0, n-1)$
\While{$Size({q\_set}) < n$}
    \For{$i:=0$ to $n-1$}
        \State $j \gets \alpha + i \;\; {mod} \;\;\; n$
        \If{$j \in {q\_set}$}
            \State \textbf{continue}
        \EndIf
        \State $T_{head} \gets Dequeue(q_j)$
        \If{$T_{head} == {NULL} $}
            \State ${q\_set} \gets {q\_set} \cup \{j\}$
            \State \textbf{continue}
        \EndIf
        \If {$ID(T_{head}) > ID(T_{new})\;\;||\;\; ID(T_{head}) \in {id\_set}$}
            \State $Enqueue(q_j, T_{head})$
            \State ${q\_set} \gets {q\_set} \cup \{j\}$
            \State \textbf{continue}
        \EndIf
        \State ${id\_set} \gets {id\_set} \cup \{ID(T_{head})\}$
        \State $p \gets M(T_{new}, T_{head})$
        \State $Enqueue(q_j, T_{head})$
        \If {$p > \theta $}
            \State  $Enqueue(q_j, T_{new})$
            \State \Return 
        \EndIf
    \EndFor
\EndWhile
\State $Enqueue(q_\alpha, T_{new})$
\end{algorithmic}
\label{algo:bfs}
\end{algorithm}

The worst case computation cost, in terms of model evaluations, for both algorithms is $O(n)$ where $n$ is the number of waiting transactions. This cost is paid by every queued transaction. 

The combination of the model evaluation and the heuristic of scheduling aborts into the same queue will tend to group high probability conflicting transactions into the same queue. We can exploit this behavior to reduce the computation cost of finding the best queue to schedule a transaction collapsing the feature vectors of the transactions in the queue to a centroid that represents the ``average'' transaction.

\subsubsection{Balanced Vector Scheduling Algorithm}
The Vector scheduler
creates a representative transaction $R_i$ for the history of transactions in each queue $q_i$. $R_i$ represents transactions in $q_i$ by averaging the feature vectors of the transactions enqueued into $q_i$. That is, the features of transaction are bit vectors but the centroid representing each queue is a vector of floats. The model $M$ remains unchanged for training purposes. The difference is in instance evaluation. To determine the best queue for a new transaction $T_{new}$ the algorithm  computes \[Enqueue(T_{new},\argmax_i = M(T_{new}, R_i))\] 
The computational cost in terms of model evaluations is $O(n)$ where $n$ is the number of queues in the system.

When we assign $T_{new}$ to $q_i$, we need to update $R_i$. This update requires latches to insure consistency of the update in the case of multiple schedulers. The results section describes the empirical cost of this contention.

The random schedule insures a balanced workload across the system. The $BFS$ scheduler somewhat balances workload due to scheduling into empty queues. The $Vector$ scheduler described above does not possess this property. To balance the workload, we additionally track the average response time for transactions coming from each queue, as well as the standard deviation. If a queue has too many transactions, the average transaction response time will increase drastically and exceed a reasonable response time limitation, which we define to be one standard deviation higher than the average response time. If the response time of queue with the highest abort probability is above this limit, the queue is blocked from receiving new transactions.  The queue is unblocked when it is empty. The new transaction is assigned to the unblocked queue with lowest response time. 

Noe that the $R$ vector is updated during an enqueue but not a dequeue. So $R$ represents a history of transactions scheduled into the queue and thus represents a centroid of the feature vectors of the history. The $Balanced$ $Vector$ $Assign$ scheduler (\cref{algo:balanced}) attempts to trade off high abort probability with a balanced workload. In the algorithm, $R[i]$ is the sum of all the vectors that have been enqueued in $q_i$ and $R_{avg}[i]$ is the average of all the vectors that have been enqueued in $q_i$. Both $R[i]$ and $R_{avg}[i]$ are zero vectors when the system initially starts. 
\begin{algorithm}
\caption{BalancedVectorAssign($T_{new}$)}
\begin{algorithmic}
\State ${max\_prob} \gets 0 $
\State ${idx} \gets -1 $
\State $V_{new} \gets Hash(T_{new})$
\For{$i:= 0$ to $n-1$}
    \State $P \gets M'(V_{new}, R_{avg}[i])$
    \If {$P \ge {max\_prob}$}
        \State  ${max\_prob} \gets P$
        \State ${idx} \gets i$
    \EndIf
\EndFor
\State ${rt} \gets GetResTimeHistory(q_{{idx}})$
\State ${rt\_avg} \gets GetResTimeAvg(rt)$
\State ${rt\_std} \gets GetResTimeStd(rt)$
\If{${rt - rt\_avg} > {rt\_std}$}
    \State ${min\_rt} \gets \infty $
    \For{$i:= 0$ to $n-1$}
        \State ${rt\_i} \gets GetResTimeHistory(q_i)$
        \If {${rt\_i} < {min\_rt}$}
            \State  ${min\_rt} \gets {rt\_i}$
            \State ${idx} \gets i$
        \EndIf
    \EndFor
\EndIf
\State $Count[{idx}] \gets Count[{idx}] + 1$
\State $R[{idx}] \gets R[{idx}] + V_{new}$
\State $R_{avg}[idx] \gets R[{idx}] / Count[{idx}]$
\State $Enqueue(q_{{idx}}, T_{new})$

\end{algorithmic}
\label{algo:balanced}
\end{algorithm}

\subsection{Supervised Scheduler Summary}
In summary, the transactions assignment execution time of BFS is $O(n)$ where $n$ is the number of queued transactions, less the early stop criterion of finding a younger transaction or finding a high abort probability transaction. The transaction execution time of Balanced Vector Assign is $O(kd)$ where $k$ is the number of queues, $d$ is the length of the feature vector (in a dense representation of floats), plus some additional cost for detecting imbalances in response times. The vector centroids are updated with each newly queued transaction (but not on transaction commit) and thus represent a history of transactions.

\section{Unsupervised Machine Learning}
The algorithms in the prior section used supervised learning to construct a model and then applied that model to the scheduling problem in various ways. An alternative is to discover clusters of transactions  that are likely to abort with other transactions in the same cluster if run concurrently. Unsupervised clustering algorithms are used to form $k$ clusters, one for each queue. When a new transaction arrives, the algorithm assigns the transaction to the cluster (queue) closest to the new transaction. Each cluster represents a group of similar types of aborts.

\subsection{Clustering Model}



Each abort instance provides a clustering example since the abort suggests clustering the two transactions, the aborted transaction and the transaction that caused the abort, into the same cluster. Our model essentially learns a clustering model from these aborts. In this model, committed transactions do not write to the log because commit information is not necessary. 

\subsubsection{Features}
Given two transactions $T_1, T_2$ such that $T_1$ aborted due to $T_2$, then $(T_1, T_2)$ is an instance of abort recorded in the log. $F(\cdot)$ extracts features from $(T_1, T_2)$ to form a feature vector as follows.
Let $V_1 = Hash(T_1) $ and $V_2 = Hash(T_2)$. Then $F(\cdot)$ is the logical \texttt{AND} of these two vectors:
\[F(T_1, T_2) = V_1 \& V_2 \]
The assumption here is that $V_1$ and $V_2$ encode the salient abort features of $T_1$  and $T_2$, respectively, and $V_1 \& V_2$ encodes evidence for the features that caused $T_1$ to be aborted. For example, suppose we set the vector size to be $8$ and we have two transactions.
\begin{align*}
    T_1 &:
    \texttt{WAREHOUSE\_ID=1} \;\; \texttt{ITEM\_ID=123} \;\; \texttt{USER\_ID=10} \\ 
    T_2 &:
    \texttt{WAREHOUSE\_ID=1} \;\; \texttt{ITEM\_ID=456} \;\; \texttt{USER\_ID=20} 
\end{align*}
Assume that the hash functions hashes the string \texttt{WAREHOUSE\_ID=1} to index 2 (starting from 0) (without collision) and suppose the results are
\begin{align*}
    Hash(T_1) = V_1 &: 00100101 \\
    Hash(T_2) = V_2 &: 10100010
\end{align*}
Then,
\begin{align*}
    F(T_1, T_2) = V_1 \& V_2 = 00100000
\end{align*}
which is a vector that represents aborts caused by sharing the attribute \texttt{WAREHOUSE\_ID=1}.

\subsubsection{Training}
Training data is gathered by observing the log while the system is running under random scheduling as in supervised learning. However, only log abort instances are logged. In particular, if a transaction aborted is by another transaction, we log
\begin{itemize}
    \item (Aborted Transaction, Conflicting Transaction)
\end{itemize}

\subsubsection{Clustering Algorithm}
We use the $k$-means scheduling algorithms with a Euclidean distance function $D(V,W)$. Although the decision boundary of k-means is linear, this algorithm converges relatively quickly in practice but more importantly can evaluate new instances quickly, on the order of O($k$) centroid comparisons. Each centroid, in a spare representation, is of length $f$, the number of features in the centroid. 
If $V = (v_1, v_2, ..., v_n)$ and $W=(w_1, w_2,...,w_n)$, then 
\[D(V,W) = \sqrt{\sum_{i=1}^n(v_i-w_i)^2}\]
Consider the following two simple clusters:
\begin{align*}
    C_1&: (0.8, 0.2) \\
    C_2&: (0.2, 0.8)  
\end{align*}
Then, transactions with $V=(1, 0)$ are clustered into $C_1$ because $D(C_1, V) < D(C_2, V)$ and those with $V=(0,1)$ are clustered into $C_2$.

Note that the clusters do not model transactions that are similar to each other in the workload, as is done in workload modeling. Consider the following two transactions:
\begin{align*}
T_1\ \ \ &\texttt{UPDATE inventory SET stock = stock-1
    WHERE}\\
    &\texttt{i\_id = 100 AND w\_id=1}\\
T_2\ \ \ &\texttt{UPDATE inventory SET stock = stock-2
    WHERE}\\
    &\texttt{i\_id = 200 AND w\_id=2}
\end{align*}
Suppose the concurrency control policy uses
row level conflict detection, then these two transactions should be in different clusters because they do not conflict. This clustering is different from workload modeling where these transactions are in the same cluster because they have similar work. Instead our clusters form different types or ``causes'' of aborts.

\subsection{Scheduling Algorithm}
After running the clustering algorithm, we obtain $k$ centroids, denoted by $C_1, C_2, ... C_k$. The next question is to design a scheduling algorithm that schedules new transactions. Given a new transaction $T_{new}$, compute the Euclidean distance between $T_{new}$ and $C_1, ... C_k$ and pick the "closest" cluster $C_c$ and assigned $T_{new}$ to queue $c$. 

The reader may notice that this method for the evaluation of an instance is an abuse of the normal machine learning framework because the training data distribution is different from the new instance data distribution. Each cluster is designed to represent a type of abort produced by a pair of transactions, but the new instance represents only a single transaction. Computing the distance between a transaction and an abort seems to be meaningless. Intuitively however, the feature vector of $T_{new}$ contains the features associated with an abort, along with some other features. Empirically, if the distance $D(V_{new}, C_i)$ is small, then $T_{new}$ is more likely to have abort type $i$. Since each cluster represents a collection of transactions, then $T_{new}$ is more likely to abort with transactions in group $i$. 

In addition, as with the balanced workload above, the balanced clustering scheduler (\cref{algo:cluster}) tracks the response time of each queue.
The $Balanced$ $Cluster$ $Assign$ scheduling algorithm performs O($k$) model instance evaluations for each new transaction that is scheduled. 
\begin{algorithm}
\caption{BalancedClusterAssign($T_{new}$)}
\begin{algorithmic}
\State $\texttt{min\_dist} \gets \infty $
\State $\texttt{idx} \gets -1 $
\For{$i:= 0$ to $n-1$}
    \State \texttt{dist} $ \gets D(V_{new}, C_i)$
    \If {$\texttt{dist} < \texttt{min\_dist}$}
        \State  $\texttt{min\_dist} \gets \texttt{dist}$
        \State $\texttt{idx} \gets i$
    \EndIf
\EndFor
\State $\texttt{rt} \gets GetResTimeHistory(q_{\texttt{idx}})$
\State $\texttt{rt\_avg} \gets GetResTimeAvg(rt)$
\State $\texttt{rt\_std} \gets GetResTimeStd(rt)$
\If{$\texttt{rt - rt\_avg} > \texttt{rt\_std}$}
    \State $\texttt{min\_rt} \gets \infty $
    \For{$i:= 0$ to $n-1$}
        \State $\texttt{rt\_i} \gets GetResTimeHistory(q_i)$
        \If {$\texttt{rt\_i} < \texttt{min\_rt}$}
            \State  $\texttt{min\_rt} \gets \texttt{rt\_i}$
            \State $\texttt{idx} \gets i$
        \EndIf
    \EndFor
\EndIf
\State $Enqueue(q_{\texttt{idx}}, T_{new})$
\end{algorithmic}
\label{algo:cluster}
\end{algorithm}

\subsection{Unsupervised Scheduler Summary}
In summary, the transactions assignment execution time of Balanced Cluster Assign is $O(kf)$ where $k$ is the number of queues, $f$ is the number of 1 features, plus some additional cost for detecting imbalances in response times. The centroids are static and thus do not have contention issues with multiple schedulers.

\section{Experimental Framework}
The experimental framework focuses on isolating external factors from the system to highlight any differences in the policies. Experiments are run on a server that is otherwise idle. The server used for experiment contains 20 cores from Intel Xeon Silver 4114 CPU @ 2.20GHz with 125GB of memory. Hyper-threading technology is not used. One thread is pinned to work-flow computations exclusively. The remaining 19 worker threads both assign and execute transactions.  In this framework, each worker thread is a scheduler and a transaction executor and the thread is dedicated to one queue of transactions. This arrangement allows us to measure the cost of scheduling algorithms but potentially introduces high contention between schedulers. The experimental design is an open queue environment where the arrival rate is used to simulate real world scenario where transactions arrive in DBMS at certain rate. Given an arrival rate, each worker thread creates and assigns enough transactions to meet the arrival rate requirements. 

All experiments are run on the \Peloton system,  an open source, stored procedure, main-memory database system.  We analyze performance {\em relative} to random scheduling, instead of absolute performance during our analysis. This type of analysis emphasizes the properties of the different schedulers, so our results generalize to other systems. Similarly, the experimental system uses optimistic concurrency control, but since the approach treats the system as a black box, the results should generalized to pessimistic concurrency control regimes. (The empirical demonstration of this claim is deferred to future work.) As part of the experiment framework implementation, for each transaction benchmark, we manually label references in the transaction workload to satisfy the canonical feature rules for the entire multistatement transaction. The steps in an experiment are as follows.

\textbf{Experimental Execution}
    \squishlist
        \item Initialize the random number generator to a new unique seed.
        \item Initialize and start the database for the benchmark.
        \item Execute transactions for 20 seconds and schedule transactions by random assignment to warm up the system.
        \item Collect transactions execution results into log file during warm-up phase in Phase I.
        \item Learn a model from a sample of the log file.
        \item Turn on measurements, then schedule and execute transactions for 300 seconds under a given policy.
        \item Wait for all transactions to finish execution.
        \item Stop the database.
    \squishend

We repeat the above three phases 3 times and report the average of measurements across the 3 experiment runs.


Transaction response time is measured  as wall clock time for a transaction from the start of the transaction assignment code until the transaction commits. So response time includes the wait time in a queue. If a transaction aborts, it is immediately restarted, and the time is included in the response time recorded for the transaction. Transaction throughput is measured as the total number of transactions to commit divided by wall clock experiment time, which does not include warm-up and learning phase. The abort rate is measured as the total number of aborts divided by the total number of attempts at execution. In particular, if a transaction aborts several times before it commits, every retry is counted. 

When a thread detects an empty queue, it checks to see if additional jobs must be generated to match the arrival rate. If not, the thread sleeps for 10 microseconds and then loops to check again. This time is charge to the measure of idle time for the thread.

Step 7 is included to insure that a large skew in the queue length is properly "charged" to our metrics, since the queue must process any remaining transactions before the experiment run is stopped. 

We use two standard OLTP benchmarks, TPC-C~\cite{TPCC} and TATP~\cite{TATP}, and one
 that models social networking website, Epinions.com~\cite{Epinions}, to evaluate the proposed methods. The benchmark parameters details are stated below. They are set to achieve relatively high abort rates.

\squishitemize
    \item \textbf{TPC-C} For TPC-C, we only include the main two types of transactions, $New Order$ and $Payment$. The number of warehouses is set to be  equal to the number of worker threads. The scale factor set to $1$. Every transaction has 50\% probability to be either $NewOrder$ or $Payment$ and the transaction warehouse id is drawn from a uniform distribution. In a $NewOrder$ transaction, the supplying warehouse for an item is selected as the home warehouse 99\% of the time and as a remote warehouse 1\% of the time. In a $Payment$ transaction, the customer is paying through the home warehouse with 85\% chance and a remote warehouse with 15\% chance.
    
    \item \textbf{TATP} The scale factor of TATP is set to $1$. The subscriber id of a transaction has a Zifian distribution with parameter $1.2$ to achieve a relatively high abort rate. 
    \item \textbf{Epinions} The Epinions workload is implemented using oltpbenchmark~\cite{Epinions} code with scale factor set to $1$. It consists of 2000 users, 1000 items and more than 7000 reviews and trusts between users. This tiny size is chosen to increase the abort rate of the benchmark. The user id and item id distributions are either uniform distributions or Zipfian distributions with parameter $1.95$, depending on the context of experiments. Zipfian distributions are used to  increase the frequency of aborts while remaining true to the realism of the benchmarks. The distribution of transaction types is shown in Table~\ref{table:epinions_txn_dist}.
    
    \begin{table}[h]
\caption{Epinions Transaction Distribution}
\begin{tabular}{lrr}
Transaction TYpe & Percentage (\%)\\ \hline \hline
GetReviewItemById & 4\\ 
GetAverageRatingByTrustedUser & 4 \\
GetItemAverageRating & 4 \\
GetItemReviewsByTrustedUser & 4\\
GetReviewsByUser  & 4 \\
UpdateItemTitle & 20 \\
UpdateReviewRating & 20 \\
UpdateTrustRating & 20 \\
UpdateUserName & 20 
\end{tabular}
\label{table:epinions_txn_dist}
\end{table}

\squishend

\section{Results and Discussion}
\label{sec:results}
We compare four schedulers (Random, BFS, Balanced Vector and Balanced k-Means Scheduler) over three benchmarks (TPC-C, TATP and Epinions). The Random Scheduler serves as the baseline for comparison. Experimentally, we step-wise increase the transaction arrival rate to the point where the maximum execution speed of the system is equal to the arrival rate, and examine the transaction throughput, response time, transactions distribution, scheduling cost and workload balance over threads.  We also analyze the underlying factors in success and failure of different scheduling algorithms. 

\subsection{Transaction Distribution Over Queues}
\label{sec:distribution}
In order to understand the transaction assigning decisions made by all these algorithms, we investigated how transactions are distributed in queues for our benchmarks.

\begin{figure}
    \centering
    TPC-C:\\
    \includegraphics[width=8cm,clip]{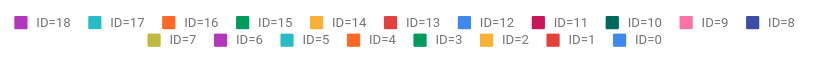}
    \includegraphics[width=6cm,clip, trim={0 0 3cm 0}]{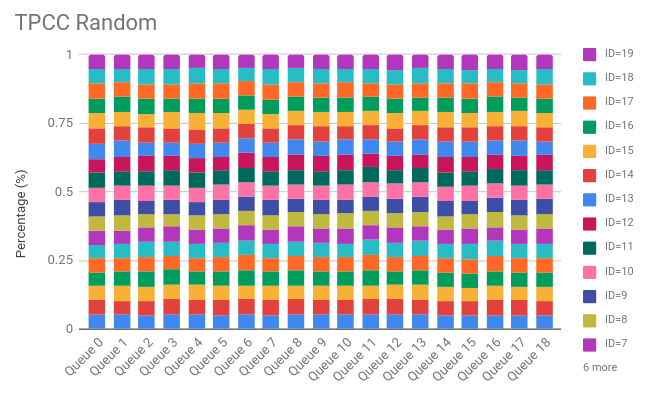}
    \includegraphics[width=6cm,clip, trim={0 0 3cm 0}]{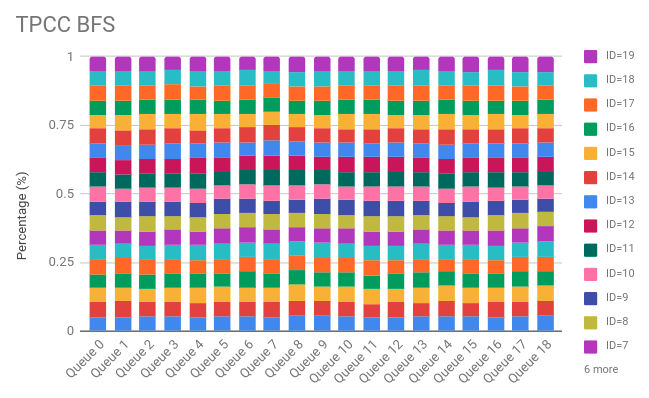}
    \includegraphics[width=6cm,clip, trim={0 0 3cm 0}]{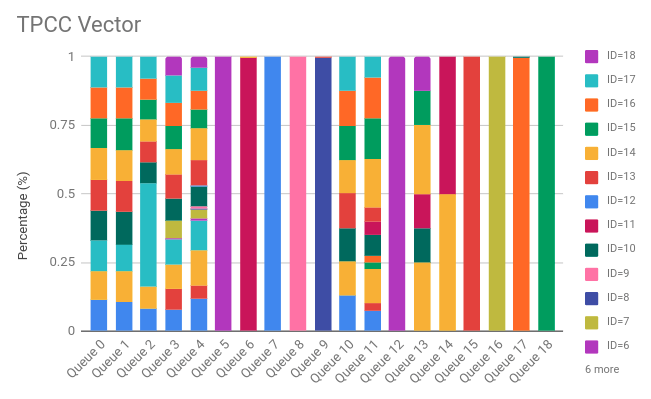}
    \includegraphics[width=6cm,clip, trim={0 0 3cm 0}]{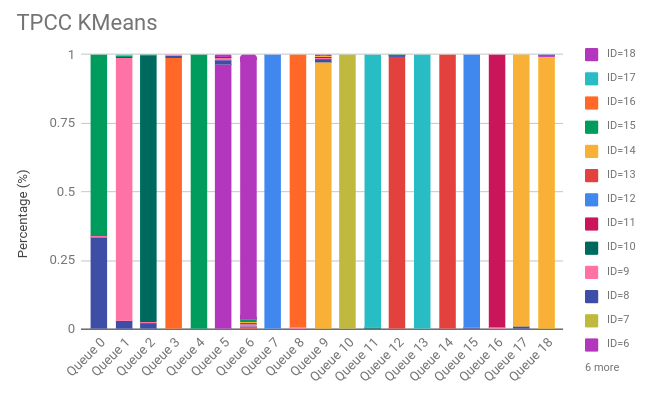}
    \caption{
        \textbf{TPC-C Transaction Distribution over Queues indexed on ID}--The ID here represents \texttt{warehouse\_id}, which has a uniform distribution. 
    }
\label{fig:tpcc_txn_dist}
\end{figure}

In the TPC-C benchmark, the transactions are uniformly assigned to warehouses (except for remote warehouses), leaving only the influence of transaction aborts. The random scheduler produces a uniform distribution of transactions when grouped by queue and then by warehouse id (\cref{fig:tpcc_txn_dist}).  BFS does not alter the distribution compared to random.  Balanced Vector does somewhat partition the transactions into queues and Balanced k-Means produces a distribution almost identical to the optimal partition of one warehouse per queue. Additional analysis of TATP and Epinions (in the appendix) reveals similar trends for those benchmarks.

\subsection{Scheduler Performance}
We ran all four schedulers at their maximum arrival rates. Throughput is measured in transaction per second (TPS) and the relative throughput is obtained by dividing by the throughput of Random Scheduler (\cref{table:bench_tp}). 
Relative throughput factors out the costs of transaction generation, assignment, and execution, and thus gives a better analysis of the relative performance of each scheduler. 

\begin{table}[h]
\caption{Throughput and Abort Rate}
\begin{tabular}{c | l p{1cm} p{1cm}}
Benchmark & Algorithm & Relative TPS & Abort Rate \\ \hline \hline
\multirow{4}{4em}{TATP} & Random & 1.000 &  0.324 \\ 
&BFS & 0.981 & 0.330\\
&Balanced Vector & 0.971 & 0.021 \\
&Balanced k-Means  & 1.544 & 0.057\\
\hline
\multirow{4}{4em}{TPCC} & Random &  1.000 &  0.534 \\ 
&BFS & 1.070 & 0.537\\
&Balanced Vector  & 0.970& 0.023  \\
&Balanced k-Means & 1.306 & 0.045\\
\hline
\multirow{4}{4em}{Epinions}&Random & 1.000 & 0.242 \\ 
&BFS &  1.103& 0.094\\
&Balanced Vector & 1.141 &  0.010\\
&Balanced k-Means  &  1.170 & 0.103\\
\end{tabular}
\label{table:bench_tp}
\end{table}



We first consider the abort rate of the two OLTP benchmarks. 
Both Balanced Vector and Balanced k-Means schedulers reduce the abort rate by more than 80\% for TATP and TPC-C. The results agree with the transaction distributions in Section 7.1.  
For standard OLTP benchmarks, BFS fails to reduce the abort rate significantly, confirming the results in the prior section. 

For the abort rate of Epinions, in Figure \ref{fig:epinions_u_txn_dist}, Balanced Vector scheduler groups transactions by their user ids, which reduces the abort rate to 0.01. However, Balanced k-Means scheduler uses 6 queues mainly for transactions with id = 0 and more than 7 queues for transactions with id = 1. Such assignment results transaction aborts and can hence only cut the abort rate by half.

For throughput for the OLTP benchmarks, 
BFS spends little time searching, but also does a poor job identifying aborts, so its throughput is similar to random.
Balanced Vector scheduler does not improve throughput even though it has the lowest abort rate because of the additional cost of comparing vectors. As described in Section 4.2.2, the Balanced Vector scheduler implementation needs to update the vector every time we assign a transaction to the queue. The cost of the work completely depends on the vector length. In our settings, the vector and length is set to 3000, and updating the vectors takes more than 80\% of the work. 


For Epinions, all the schedulers outperform random since the abort rate has a much higher impact on throughput due to longer running aggregation queries included in the benchmark. The performance improvement is roughly in order of the reduction in abort rate. Balanced k-Means outperforms Balanced Vector even with a higher abort rate due to the lack of update costs during scheduling. In Balanced k-Means, the centroids are static and read only.

Note that in a system where aborts are more costly, Balanced Vector can potentially outperform Balanced k-Means. We plan to explore this issue in future work.

\subsection{Response Time vs.~Throughput}
\begin{figure}[t!]
    \centering
    TPC-C:
    \includegraphics[width=8cm,clip]{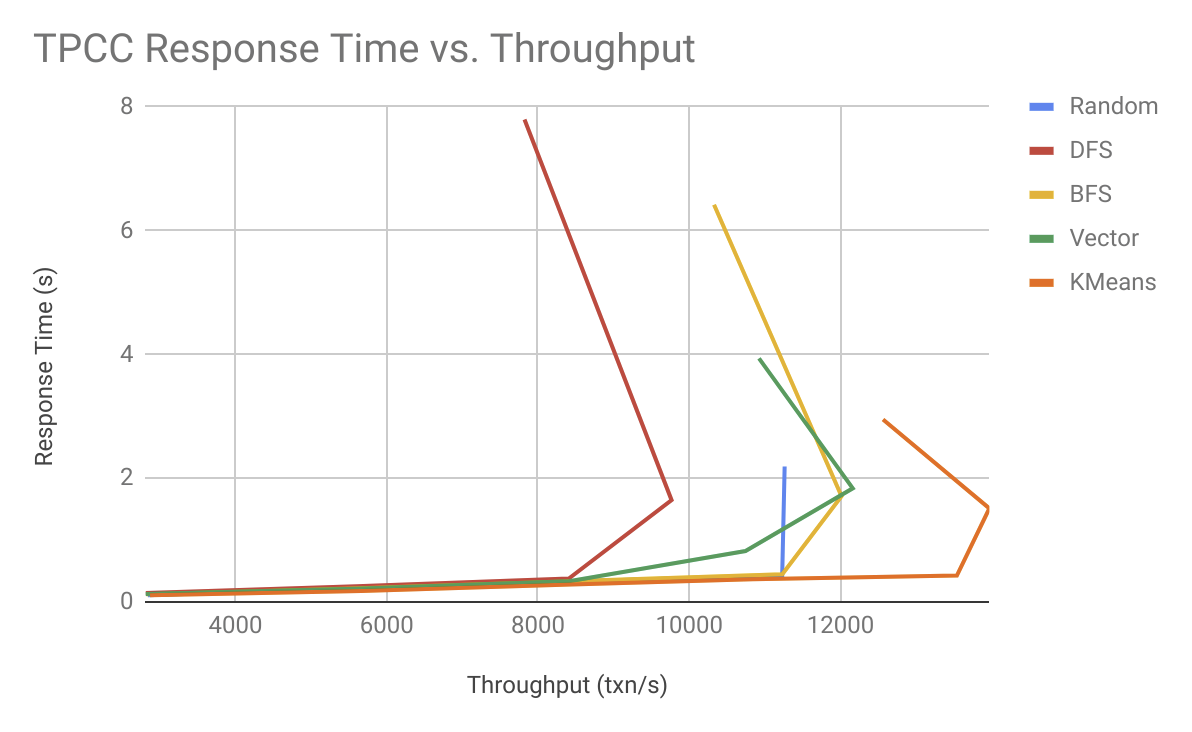}
    TATP:
    \includegraphics[width=8cm,clip]{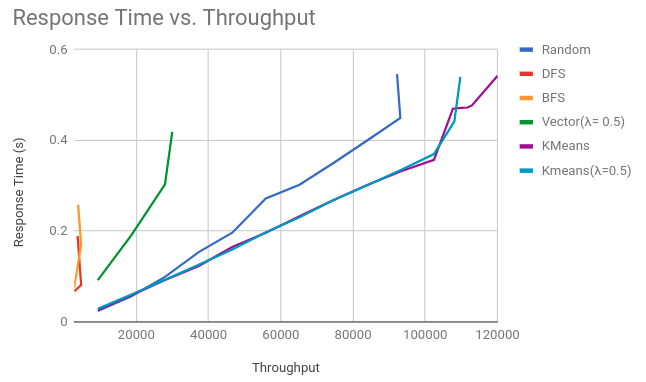}
    Epinions:
    \includegraphics[width=8cm,clip]{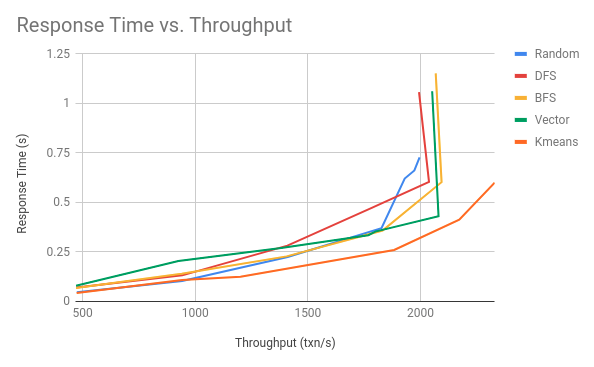}
    \caption{
        \textbf{Response Time vs.~Throughput} -- Each point in the graph represents an experiment run that generates a (throughput, response time) point. Lines connect the same scheduler as arrival time increases. The spikes in the graph correspond to the arrival rate exceeding the processing rate of the system for the given scheduler.
    }
\label{fig:rt_vs_tp}
\end{figure}

We examine the response time and throughput of all five schedulers by gradually increasing the arrival rate until the system reaches or exceeds its maximum execution speed (\cref{fig:rt_vs_tp}).  The fact that throughput becomes worse when the arrival rate is too high is because every worker thread is also a scheduler and it is forced to assign enough transactions to meet arrival rate requirement. If the arrival rate is too high, then more work is put on transaction assignment instead of execution. 

The Balanced k-Means scheduler is the winner of these five algorithms. It has a higher maximum throughput and at a given arrival rate, k-Mean scheduler cuts the response time by about 10\%. 

\subsection{Workload Skew}
The analysis of \ref{sec:distribution} reveals the proportion of workload intra-queue, but does not analysis inter-queue differences in workload. Workload balance is another important feature we want in our schedulers. A more balanced workload means more cores utilization. 
To examine the workload skew, measuring queue length ignores the work involved in individual transactions in a queue. Instead, we measure queue idle time. We lowered the arrival rate for each benchmark to a point where all the schedulers were operating effectively. We fixed the arrival rate for all schedulers for the given benchmark at this arrival rate and measured the total idle time for each thread. To measure workload skew we  compute the standard deviation among thread idles times for all four schedulers and all three benchmarks.

The results in Table \ref{tab:idle_time} show that random scheduler has the lowest standard deviation, as expected. It means that the work is evenly distributed among threads. Balanced Vector scheduler has a relatively high standard deviation. This skew implies that Balanced Vector scheduler sacrifices workload balance for abort rate. The Balanced k-Means scheduler also maintains a relatively low standard deviation, meaning the workload is also not very skewed. Thus Balanced k-Means more effectively balances the trade-off between concurrency and abort rates, leading to improved performance.
\begin{table}[h]
\begin{center}
\begin{tabular}{ c  c  c  c}
Scheduler & TPC-C & TATP & Epinions \\ \hline \hline
Random & 0.030 & 0.077 & 0.090 \\
BFS & 0.045 & 0.073 & 0.099 \\
Balanced Vector & 0.189 & 0.714 & 0.768 \\
Balanced k-Means & 0.050 & 0.204 & 0.213 \\
\hline
Arrival Rate (tps) & 10,000 & 30,000 & 2,000
\end{tabular}
\caption{Standard deviation of the total wall clock idle time available for each queue. Higher standard deviation indicates higher skew in the work performed by the queues. For each benchmark, the arrival rate is set to be close to the maximum arrival rate for the slowest scheduler for that benchmark. This arrival rate allows direct comparisons between schedulers for a benchmark. } 
\label{tab:idle_time}
\end{center}
\end{table}

\subsection{Balancing Response Time}

\begin{table}[h]
\begin{center}
\begin{tabular}{c|c|c|c}
Scheduler  & TPC-C & TATP & Epinions\\ \hline \hline
(unbalanced) k-Means &  0.27 & 0.42 & 1.3 \\
Balanced k-Means & 0.26 &0.27 & 0.47  \\
\hline
Arrival Rate (tps) & 10,000 & 30,000 & 2,000
\end{tabular}
\caption{Balanced and Unbalanced k-Means -- Average response time in seconds.}
\label{tab:balance}
\end{center}
\end{table}

Balanced k-Means contains additional work to detect imbalanced in response time. Additional experiments (Table~\ref{tab:balance}) compared balanced k-Means with a k-Means scheduler without the balancing code (unbalanced). 
This data confirms that response time balancing is critical to an effective intelligent scheduler because response times are significantly reduced in two of the three benchmarks.


\subsection{Adaption and Stability of Balanced k-Means}

\begin{table}
\begin{tabular}{r r r r}
Round & TPCC & TATP & Epinions \\ \hline \hline
0 & 1.00 & 1.00	& 1.00 \\
1 & 1.73 &	1.42 &	1.17 \\
2 &	1.71 &	1.53 &	1.18 \\
3 &	1.72 &	1.46 &	1.24 \\
4 &	1.72 &	1.46 &	1.18 \\
5 &	1.73 &	1.42 &	1.19 \\
6 &	1.74 &	1.52 &	1.20 \\
7 &	1.74 &	1.43 &	1.19 \\
8 &	1.71 &	1.45 &	1.20 \\
9 &	1.73 &	1.46 &	1.18 \\ \hline
Mean $\overline{x}$ &	1.73 &	1.46 &	1.19 \\
Standard Deviation $\sigma$ &	0.011 &	0.037 &	0.018 \\
Confidence Interval $\pm$ &	0.007 &	0.024 &	0.012 \\ \hline
\end{tabular}
\caption{Relative deviation of throughput for each round of model construction. The system warms up by using random scheduling (row 0). Each subsequent round (row 1-9) discards the prior model and builds a new model constructed from a sample of the log the previous run of the system for 30 seconds. The rows 0-9 list the relative throughput compared to row 0 for each benchmark. Row $\overline{x}$ lists the mean for rows 1-9. Row $\sigma$ lists the standard deviation for rows 1-9. Row $\pm$ lists the confidence interval at 95\% for rows 1-9. The flat performance across rounds demonstrates that the system is stable.}
\label{tab:deviation}
\end{table}

The system must adjust to changes in the workload. The system proceeds in rounds of 20 seconds. After 20 seconds, the system discards its model and builds a new model based on a random sample of 500 lines the prior round's log records. This arrangement means that the system is never more than 20 seconds behind the latest distribution. Because building the k-Means model is cheap, this simple design is sufficient for the system to adapt to any workload. 
This design may suffer from feedback problems however, because the input of the prior round is fed into the next round, a common issue in machine learning. We ran an experiment to check for this problem (Table~\ref{tab:deviation}) and determined that the system is remarkably stable. Essentially only minor variations in performance are observed from one round to the next.

\section{Related Work}

\label{sec:related}

Several recent works improve the performance of locking system in main memory databases~\cite{ren2012lightweight,tian2018contention}. This line of work is complementary to our approach. Overall our work is most closely related to the general area of self driving databases~\cite{Pavlo:2017}.

\textbf{Partitioning Data}.
One  technique to deal with contention
partitions the data to reduce aborts \cite{Stonebraker:2007:EAE:1325851.1325981, Thomson:2012:SIGMOD, weldon2013data}. Adopting partitioning
as a core assumption, along with several additional design
decisions, improves performance. However,  performance problems still remain\cite{Yu:2014:SAE:2735508.2735511} and partitioning is problematic because some database schemas are not easily
partitionable \cite{Curino:2010:SWA:1920841.1920853,Pavlo:2012:SAD:2213836.2213844,tian2018contention,prasaad-2018}. 
In an informal sense, partitioning is data centric since the partitioning is generally based on the actual read/write sets of transactions. In contrast our work is transaction centric since it focuses on the read/write set predicted from the transaction statement without direct reference to the data.

Another OLTP architectural technique to avoid aborts focuses on careful
implementation of ACID transaction properties to achieve improved
performance \cite{kim2016ermia,Tu:2013:STM:2517349.2522713,yu2016tictoc}. Systems based on this approach offer improved performance over all database schemas and do
not suffer from the constraints of partitioning. However, these systems still suffer under high contention workloads.

\textbf{Semantic analysis}.
A deterministic approach to the analysis of code is followed in~\cite{Manjhi:2009}. The approach merges multiple database calls in an application code loop back into a single SQL query. The approach identifies dependencies between database calls and extracts the independent statements to be run in parallel to decrease latency. A similar approach, extended to use optimization, is followed in QURO~\cite{Yan:2016:VLDB}, to reorder statements to reduce transaction conflict under a two-phase locking protocol. 
However, reordering requires complex analysis for query dependency. 

Deterministic analysis of transactions to detect aborts \cite{Bailis:2014:VLDB, Thomson:2012:SIGMOD} has the advantage of deriving deeper precise knowledge of the transactions structure at a higher cost of analysis. 
In contrast, we choose a lightweight, imprecise analysis to detect evidence related to transaction aborts. 


Zhang et al.~\cite{ZhangICDE2018} explores scheduling in main memory databases by considering different lightweight analysis of transactions and using a static representation for scheduling. We borrowed the notion of canonical representations of attributes from this work. In contrast, we use machine learning algorithms to learn distributions of aborts. In effect our results show that a machine learning algorithm can learn distributions to produce more intelligent and better performing schedulers.

In~\cite{Harchol-Balter:2003}, the performance impact of scheduling based on the size of static HTTP requests to a server is investigated. 
The paper improves server response time by scheduling draining network socket buffers according to Shortest Remaining Processing Time queuing policy instead of Round Robin or random. 
Priorities on the queues are set by the number of bytes left to read from a static file.
The paper and our work both use scheduling to improve performance based on a feature that characterizes the work remaining to be done for a an operation.
The main difference is that the feature is a property of the transaction in the paper, where as in our work the feature is a signal generated by the transaction processing system.

Ic3~\cite{Wang:2016:SMD:2882903.2882934,Shasha:1995:TCA:211414.211427} constructs a dependency graph and maintains the dependency graph for running transactions to provide efficient serializability. The main similarity between this work and our approach is that both analyze dependencies transactions to improve performance. However, our dependency structure is implied in the statistical relationship between references that occur in transactions, as opposed to an explicit dependency model.

\textbf{Scheduling}.
Some researchers utilize hardware to accelerate transaction processing~\cite{hardware1,hardware2,hardware3,hardware4,hardware5}. But they do not address contention issue using scheduling methods. In \cite{Blake:2009}, scheduling for hardware transactional memory is studied. Transaction boundaries in this context are annotated in source code. The program counter is used as the transaction identifier. The paper describes a scheduling system that maintains a global data structure that records a history of aborts between transactions and information about each transaction.
When a processor is ready to schedule a thread, it examines the structure and decides to execute, stall briefly, or swap in a new thread. 
In addition the paper computes a metric of a pair-wise probability of abort between two transaction. Although direct contrasts between the two works are difficult, the results suggest that simple global counting methods should be studied as an alternative to probabilistic models.

One area of application program analysis focuses on improvement performance (network latency in particular) through heuristic program analysis~\cite{Manjhi:2009}. The work is complementary to ours since it is focused on transformation of the application to improve performance as opposed to our scheduling approach.

\textbf{Data and functionality partitioning}. In H-Store~\cite{Stonebraker:2007:EAE:1325851.1325981}, a stored procedure main memory database system, the database is chopped into partitions and a thread controls a partition at any point in time. The transactions execute serially for a given partition. HyPer~\cite{hyper:icde11} also processes transactions using a single-thread. It supports both OLTP and OLAP workloads (the latest version of Hyper uses optimistic multi-versioning~\cite{hyper:sigmod15}). In DORA~\cite{Pandis:2010:DTE:1920841.1920959}, a transaction is split  into sub-transactions and a sub-transaction is assigned to a single partition to execute. 

Cheung et al.~propose a technique of automatic code partitioning using program analysis~\cite{Cheung:2012:APD:2350229.2350262}. By code partitioning they mean the partitioning of the application program between the application and stored procedures on the DBMS. The approach performs a program analysis of the application and then chops up the program into pieces with respect to network latency. Some pieces are executed on the application server and other pieces are executed on the database server. In effect, store procedures are automatically defined through program analysis. 

\textbf{Multi-core optimization}. Larson et al.~propose an optimistic validation protocol that avoids using global critical section~\cite{Larson:2011:HCC:2095686.2095689}. Their methods address the scalability problem for both pessimistic and optimistic concurrency control. In Silo~\cite{Tu:2013:STM:2517349.2522713}, an optimistic in-memory multi-core DBMS is proposed to avoid centralized contention points. It provides a commit protocol with serializability and eliminates all shared-memory writes. Johnson et al. use speculative lock inheritance to decrease the frequency of contended latch acquisitions~\cite{Johnson:2009:IOS:1687627.1687682}. Their method passes contended logical locks without invoking the lock manager. 

ORTHRUS~\cite{Ren:2016:DPS:2882903.2882958} separates execution threads from concurrency control threads. Execution threads send message to concurrency control threads to acquire locks for accessed records. If a transaction accesses several records, the execution thread might send message to different concurrency control threads. In order to fully utilize CPU cycles, concurrency control threads may send message to other concurrency control threads to help acquiring locks. 
All of these systems focus the concurrency control protocol design and thus are complimentary to our approach. Our approach builds another layer that is on top of concurrency control and can be used by any concurrency control protocol. 

Partitioning of batches transactions of transactions is closely related to the approach in this paper. In \cite{prasaad-2018}, the read-write set of transactions is used to construct clusters of transactions which are then executed in parallel without concurrency control, with a few residual transactions executed with normal concurrency control (or sequentially).  In \cite{mu2014extracting}, transaction chopping is used to determine conflicting parts of transactions which are then split (or merged) into blocks that are executed on a distributed H-store style system. In principle, both these works would benefit from our approach, since a k-Means clustering would reduce the size of the sets of transactions considered by both works. 
\section{Conclusions}

In this paper, we conjectured that (i) the patterns of transaction aborts could be learned through the use of machine learning models, and (ii) those models could be used to improve the scheduling of transactions to reduce aborts and increase throughput. The paper systematically explored both supervised and unsupervised models for machine learning to create intelligent schedulers. 
In summary, the paper provides preliminary evidence that improved performance via intelligent scheduling is indeed possible in main memory database systems.



\bibliographystyle{ACM-Reference-Format}
\bibliography{ref}

\appendix

\section{Appendix}
\begin{figure}
    \centering
    TATP:\\
     \includegraphics[width=8cm,clip]{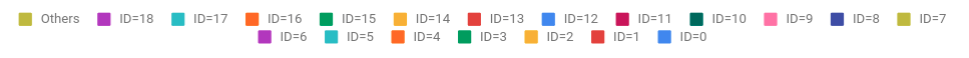}
    \includegraphics[width=6cm,clip, trim={0 0 3cm 0}]{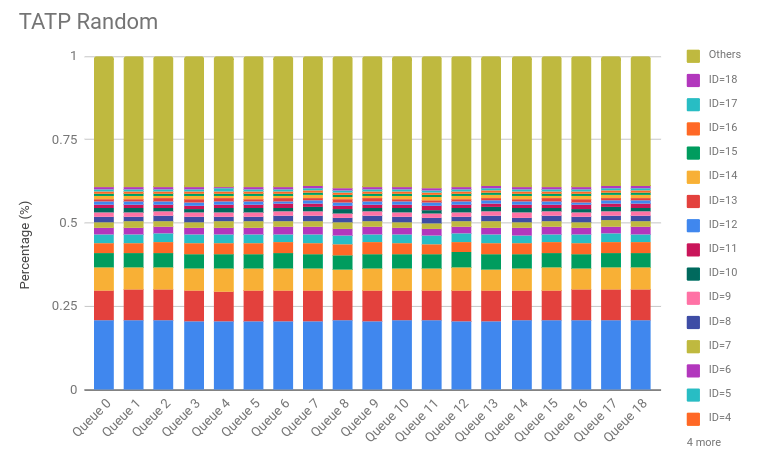}
    \includegraphics[width=6cm,clip, trim={0 0 3cm 0}]{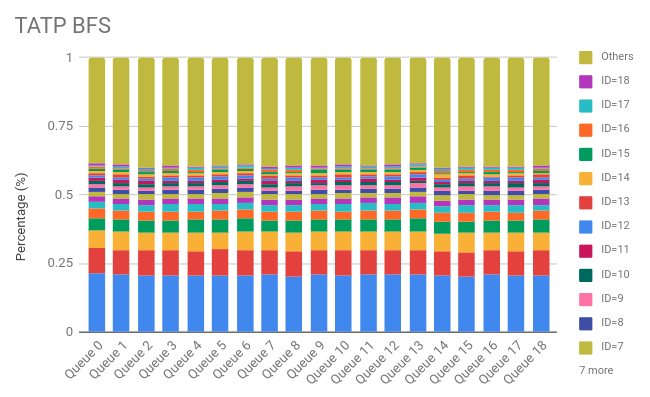}
    \includegraphics[width=6cm,clip, trim={0 0 3cm 0}]{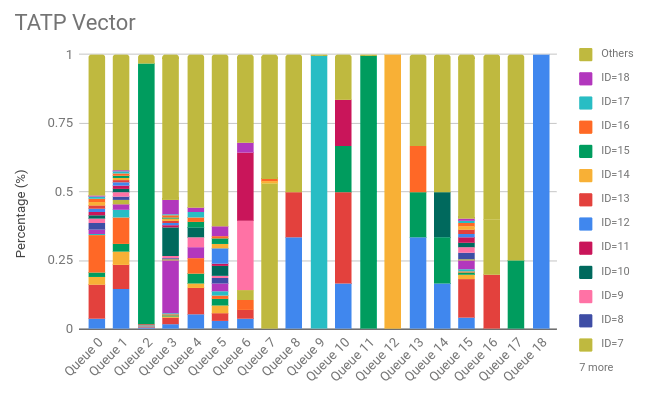}
    \includegraphics[width=6cm,clip, trim={0 0 3cm 0}]{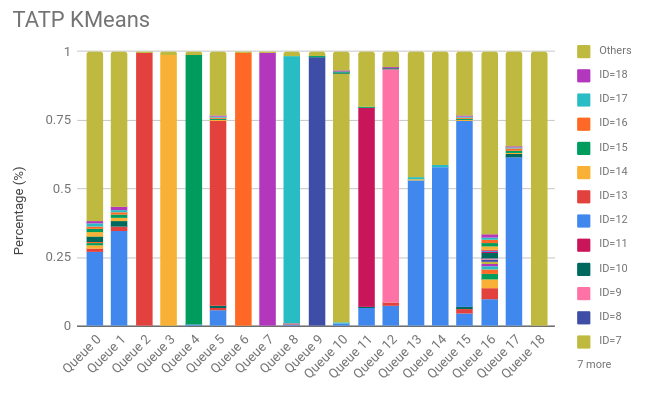}
    \caption{
        \textbf{TATP Transaction Distribution over Queues indexed on ID}-The ID here represents \texttt{subscriber\_id}, which has a Zipfian distribution. We categorize the first 19 IDs and the Other category includes the remaining ids. 
    }
\label{fig:tatp_txn_dist}
\end{figure}

\begin{figure}
    \centering
    Epinions U\_ID:\\
    \includegraphics[width=8cm,clip]{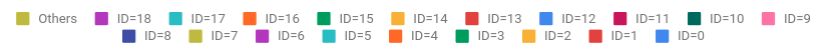}
    \includegraphics[width=6cm,clip, trim={0 0 3cm 0}]{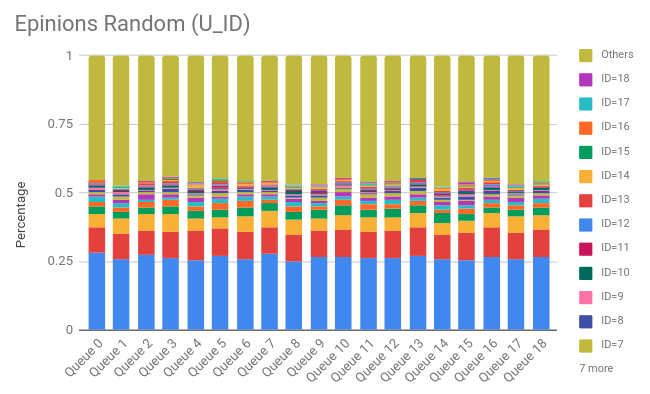}
    \includegraphics[width=6cm,clip, trim={0 0 3cm 0}]{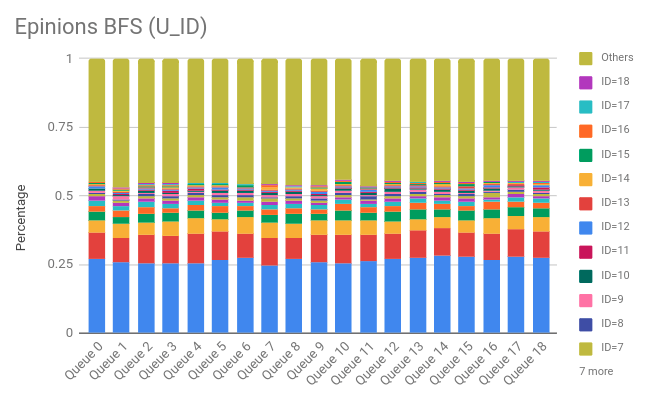}
    \includegraphics[width=6cm,clip, trim={0 0 3cm 0}]{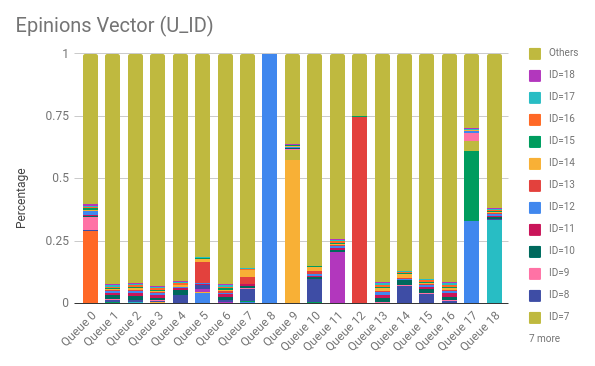}
    \includegraphics[width=6cm,clip, trim={0 0 3cm 0}]{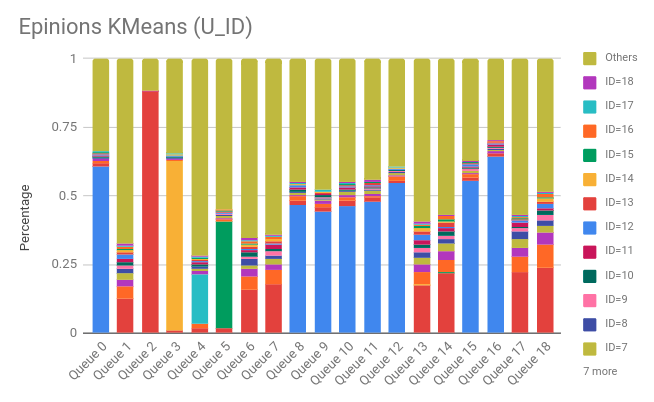}
    \caption{
        \textbf{Epinions Transaction Distribution over Queues indexed on ID}-The ID here represents \texttt{user\_id}, which has a Zipfian distribution. We categorize the first 19 IDs and the Other category includes the remaining ids. 
    }
\label{fig:epinions_u_txn_dist}
\end{figure}

For TATP, We group transactions first by queue and then by their \texttt{SUBSCRIBER\_ID} to see how they are proportionally distributed (\cref{fig:tatp_txn_dist}). For TATP benchmark, the random scheduler places transactions uniformly over queues, thus the proportion of execution for each queue of each transaction grouped by subscriber id is nearly the same. The height of the colored bar indicates the proportion of execution of a particular identifier in the queue. The Zipfian skew of the benchmark is clearly visible.


The Balanced Vector and k-Means schedulers tend to dedicate queues to transactions with certain \texttt{SUBSCRIBER\_ID} range. Since transactions with the same ids typically conflict with each other when run concurrently, both the Balanced Vector and k-Means schedulers attempt to reduce conflicts by avoiding running transactions with the same id concurrently. 

Epinions is a more complex benchmark in the sense that some transactions aggregate over either user ids or item ids. We skewed the distribution of user ids and kept the distribution of item ids uniform. The same analysis (\cref{fig:epinions_u_txn_dist}) shows that for random scheduling, user id are uniformly spread across queues and that within a queue there is high skew in the user id. Balanced Vector adjusted more for transaction conflicts and Balanced k-Means does the best job of adjusting. An analysis of transaction distributions by item ids indicates transactions are uniformly distributed across all queues. We have also tested keeping user ids uniform and skewing only item ids and both Balanced Vector and Balanced k-Means can learn to distribution of transactions accordingly. This test reveals that the system properly learns distributions of data that cause conflicts across a workload. 
\end{document}